\documentclass[aps,prd,showpacs,amsmath,amssymb]{revtex4}

\begin{document}

\title{ Hypercolor tower}
\rightline{\small ITEP-LAT/2009-02} \vskip 30mm
{
\vspace{1cm}

\author{\firstname{M.A.}~\surname{Zubkov}}
\affiliation{\rm ITEP, B.Cheremushkinskaya 25, Moscow, 117259, Russia}
\affiliation{\rm Moscow Institute of Physics and Technology, 141700,
Dolgoprudnyi, Moscow Region, Russia.}

\begin{abstract}
We construct the ultraviolet completion of the Standard Model that contains an
infinite sequence of Hypercolor gauge groups. So, the whole gauge group of the
theory is $... \otimes SU(5)\otimes SU(4) \otimes SU(3) \otimes SU(2) \otimes
U(1)$. Here $SU(4)$ is the Technicolor group of Farhi - Susskind model. The
breakdown of chiral symmetry due to the the Technicolor gives rise to finite
$W$ and $Z$ boson masses in a usual way. The other Hypercolor groups are not
confining. We suggest the hypothesis that the fermion masses are not related in
any way to technicolor gauge group. We suppose that the fermion mass formation
mechanism is related to the energies much higher than the technicolor scale.
Formally the fermion masses appear in our model as an external input. In the
construction of the theory we use essentially the requirement that it posseses
an additional discrete symmetry $\cal Z$ that is the continuation of the $Z_6$
symmetry of the Standard Model. It has been found that there exists such a
choice of the hypercharges of the fermions that the chiral anomaly is absent
while the symmetry $\cal Z$ is preserved. 
\end{abstract}

\pacs{12.15.-y, 11.15.Ha, 12.10.Dm}

\maketitle

\section{Introduction}

Thinking about the possible ultraviolet completion of the Standard Model, we
encounter Technicolor and Extended Technicolor theories, Little Higgs models,
supersymmetry, extra dimensions, and Tev - scale gravity (see, for example,
\cite{Technicolor,ExtendedTechnicolor,LittleHiggs,SUSYLHC,Extra,TevGrav}).
However, basing on the present data we cannot make a definite choice. Probably,
the data of LHC coming soon will help more.

It is worth mentioning that the Standard Model itself cannot describe physics
at the energies above $1$ Tev. The conventional way to explain this is based on
the concept of "naturalness" and is related to the treatment of the fine tuning
of Higgs sector mass parameter as unnatural \cite{TEV}. Besides, it was shown
recently, that the Standard Model in lattice regularization cannot have in
principle the value of the ultraviolet cutoff larger than about $1$ Tev
\cite{VZ2008,BVZ2007}.

In this paper we suggest the model that is based on the ideas of Technicolor.
However, our approach differs essentially from the conventional Technicolor
approach. The main difference is that we do not incorporate to the theory the
Extended Technicolor interactions. That's why the Technicolor interaction
serves only as a source of Electroweak gauge symmetry breaking. As for the
fermion masses, we suggest the following hypothesis: {\it the source of the
fermion masses is not related to Technicolor interactions and to the
transformation of the technifermions to the other physical states}. We suppose
that the fermion mass formation mechanism may be related to the energies much
higher than it is usually implied in the Extended Technicolor models.  The
second important supposition of our paper is that the ultraviolet completion of
the Standard Model contains an infinite sequence of the gauge groups $...
\otimes SU(5)\otimes SU(4) \otimes SU(3) \otimes SU(2) \otimes U(1)$. We call
this sequence {\it Hypercolor Tower}, the groups $SU(N)$ of this sequence for
$N>3$ are called hypercolor groups. The Electroweak symmetry breaking occurs
due to the $SU(4)$ subgroup. That's why we refer to $SU(4)$ as to the
Technicolor group.

Thus we refuse to accept the conventional  Extended Technicolor scheme. In
spite of this below we briefly remind this scheme and its difficulties. For the
review see \cite{Technicolor,ExtendedTechnicolor} and references therein. In
the Technicolor theory the new Nonabelian gauge interaction is added with the
scale $\Lambda_{TC} \sim 1$ Tev, where $\Lambda_{TC}$ is the analogue of
$\Lambda_{QCD}$. This new interaction is called Technicolor. The correspondent
new fermions are called technifermions. The Electroweak gauge group acts on the
technifermions. Therefore, breaking of the chiral symmetry in Technicolor
theory causes Electroweak symmetry breaking. This makes three of the four
Electroweak gauge bosons massive. However, pure Technicolor theory cannot
explain the appearance of fermion masses.

Usually in order to make Standard Model fermions massive extra gauge
interaction is added, which is called Extended Technicolor (ETC)
\cite{Technicolor,ExtendedTechnicolor}. In this gauge theory the Standard Model
fermions and technifermions enter the same representation of the Extended
Technicolor group. Standard Model fermions become massive because they may be
transformed into technifermions with ejecting of the new massive gauge bosons.
Then the quark and lepton masses are evaluated at one loop level as
$m_{q,l}\sim \frac{N_{TC}\Lambda_{TC}^3}{\Lambda_{ETC}^2}$, where
$\Lambda_{TC}$ is the Technicolor scale while $\Lambda_{ETC}$ is the scale of
the new strong interaction called Extended Technicolor. (Spontaneous breakdown
of Extended Technicolor symmetry gives rise to the mass of the new gauge bosons
 of the order of $\Lambda_{ETC}$.)

The ETC models suffer from extremely large flavor - changing amplitudes and
unphysically large contributions to the Electroweak polarization operators
\cite{Technicolor}. The possible way to overcome these problems is related to
the behavior of chiral gauge theories at large number of fermions or for the
higher order representations. Namely, the near conformal behavior of the
Technicolor model allows to suppress dangerous flavor changing currents as well
as to decrease the contribution to S - parameter
\cite{Appelquist,minimal_walking}. However, the generation of $t$ - quark mass
in these models still causes serious problems\footnote{Nevertheless, see
\cite{Sannino_t}, where the way to solve the problem with the $t$ - quark is
suggested.}.

As it was already mentioned, in the present paper we avoid the mentioned
problems specific for the ETC models because we do not require that the fermion
masses are related in any way to Technicolor interactions. We suppose, that the
chiral symmetry breaking in the Technicolor theory gives rise to the gauge
boson masses only. The formation of fermion masses remains out of our model. We
only notice here that the fermion masses in relativistic theory is related to
the transition amplitude between the right handed and left handed fermions.
That's why any process that leads to appearance of such amplitude may be
treated as the fermion mass formation mechanism. In particular, we may suppose,
that the processes like this happen at extremely high energies, probably, even
of the order of Plank mass. So, formation of fermion masses may, in principle,
be related to quantum gravity.

In order to incorporate formally fermion masses to the chiral invariant theory
we introduce the auxiliary field $\Omega \in SU(2)$. We imply that there is no
dynamical term in the action that contains its derivatives. The physical sense
of this field is that it peeks up the parity partner for each right - handed
spinor. At the same time the theory possesses chiral invariance at the level of
bare action. In a certain sense $\Omega$ plays the role of the usual Higgs
field with frozen radius and without dynamical term in the action. The gauge
group of the theory is chosen to be the infinite product of $SU(N)$ groups and
the gauge group of $SU(4) \otimes SU(3) \otimes SU(2) \otimes U(1)/Z_{12}$
Farhi - Susskind Technicolor theory \cite{FS}.

In order to fix the hypercharge assignment of the model we require that the
theory is invariant under the additional discrete $\cal Z$ symmetry. This
symmetry is the continuation of the $Z_6$ symmetry of the Standard Model
\cite{Z6,Z6f,BVZ2003,Z2007} to the Hypercolor models \cite{Z2007t}. It has been
found long time ago, that the spontaneous breakdown of $SU(5)$ symmetry in
Grand Unified Theory actually leads to the gauge group $SU(3)\times SU(2)
\times U(1)/Z_6$ instead of the conventional $SU(3)\times SU(2) \times U(1)$
(see, for example, \cite{Z6} and references therein). However, the $Z_6$
symmetry is not the subject of the $SU(5)$ unification only. Actually, the
$Z_6$ symmetry is present in the Standard Model itself without any relation to
the particular Unified theory \cite{BVZ2003,Z2007,Z6f}. The $Z_6$ symmetry is
rather restrictive and it forbids, for example, the appearance of such
particles as left - handed Standard Model fermions with zero hypercharge. It
was shown in \cite{BVZ2003}, that the Unified models based on the Pati - Salam
scheme \cite{PATI} may possess the $Z_6$ symmetry. Besides, it was found that
in the so - called Petite Unification models (also based on the Pati-Salam
scheme) the additional discrete symmetry is present ($Z_2$ or $Z_3$ depending
on the choice of the model)  \cite{Z2007}.

The reason of the application of this symmetry to our construction is that we
guess the $Z_6$ symmetry of the Standard Model is not accidental. That's why,
we suppose it must emerge in a certain way in the more fundamental theory.
Besides, we find that the $\cal Z$ symmetry has a certain influence on the
monopole content of the hypothetical Unified theory that incorporates our
Hypercolor tower as a low energy approximation.

The paper is organized as follows. In the $2$ - nd section we describe the
basic ingredients of our model, i.e. the gauge group and the sequence of
fermions. In the third section we introduce parity conjugation of two -
component spinors used in our model to incorporate fermion masses. In the $4$ -
th section we describe the $Z_6$ symmetry of the Standard Model and the chosen
way to continue it to the Hypercolor groups. In the $5$ - th section we
describe the first element of the sequence of Hypercolor groups, i.e. the Farhi
- Susskind Technicolor $SU(4)$ interactions. We explain how the $\cal Z$
symmetry fixes the hypercharge assignment for technifermions. In the $6$ - th
section the way to introduce fermion masses to the theory is described. In the
$7$ - th section the formation of chiral condensates in our model is described.
In the $8$ - th section we describe the next element in the sequence of
Hypercolor groups, i.e. the $SU(5)$ interactions. In the $9$ - th section the
generalization of our consideration to the Hypercolor groups $SU(N)$ with
arbitrary $N$ is explained. In the $10$ - th section we discuss the relation
between the $\cal Z$ symmetry and the properties of the hypothetical Unified
theory.  In the $11$ - th section the dynamics of Hypercolor interactions is
briefly reviewed. In the $12$ - th section we end with our conclusions.

\section{The basic ingredients of the model}

In our approach the theory contains $U(1)$ gauge group and the groups $SU(N)$
with any $N$. So, the gauge group of the theory is

\begin{equation}
G = ... \otimes SU(5)\otimes SU(4) \otimes SU(3) \otimes SU(2) \otimes
U(1)/{\cal Z}, \label{G}
\end{equation}
where $\cal Z$ is the discrete group to be specified below.

Next, we suppose, that in the theory any fermions are present that belong to
the fundamental representations of the $SU(N)$ subgroups of $G$. So, the
possible fermions are right - handed $\Psi_{A, Y}^{\alpha i_{k_N} ... i_{k_3}
i_{k_2}}$ and left - handed $\Theta_{\dot{\beta} A, Y}^{ i_{k_N} ... i_{k_3}
i_{k_2}}$, where $\alpha$ and $\dot{\beta}$ are spinor indices, $A$ enumerates
generations while index $i_k$ belongs to the subgroup $SU(k)$. Here $Y$ is the
$U(1)$ charge of the given fermion. In particular, the fermions $\Psi_{A; Y}$
are present that have no indices and the only subgroup that acts on $\Psi_{A;
Y}$ is $U(1)$. Moreover, we suppose that the fermions are present such that $G$
does not act on them at all. We denote them $\Psi_{A;0}$. All fermions in the
theory are two - component spinors. We also suppose from the very beginning
that the $SU(2)$ group acts on the left - handed spinors only. The action of
parity conjugation on them will be considered later. For the simplicity we omit
below both spinor and generation indices. So, our fermions are

\begin{eqnarray}
U(1): && \Psi_0, \Psi_{Y_{1}}, \Psi_{Y^{\prime}_{1}},...;\nonumber\\
U(1), SU(2): && \Theta^{i_2}_{Y_2}, \Theta^{i_2}_{Y^{\prime}_2}, ...; \nonumber\\
U(1), SU(3): && \Psi^{i_3}_{Y_3}; \Psi^{i_3}_{Y^{\prime}_3}, ...; \nonumber\\
U(1), SU(2), SU(3): && \Theta^{i_3 i_2}_{Y_{32}},\Theta^{i_3 i_2}_{Y^{\prime}_{32}},... ;\nonumber\\
U(1), SU(4): && \Psi^{i_4}_{Y_4},\Psi^{i_4}_{Y^{\prime}_4},...;\nonumber\\
U(1), SU(2), SU(4): && \Theta^{i_4 i_2}_{Y_{42}},\Theta^{i_4 i_2}_{Y^{\prime}_{42}},...;\nonumber\\
U(1), SU(3), SU(4): && \Psi^{i_4 i_3}_{Y_{43}}, \Psi^{i_4 i_3}_{Y^{\prime}_{43}},...;\nonumber\\
U(1), SU(2),SU(3),SU(4): &&  \Theta^{i_4 i_3 i_2}_{Y_{432}},\Theta^{i_4 i_3
i_2}_{Y^{\prime}_{432}},...; \nonumber\\
&&...\label{F}
\end{eqnarray}
Here in each row we list the subgroups of $G$ that act on the fermions listed
in the row. In each row the allowed values of $U(1)$ charge are denoted by $Y,
Y^{\prime}$, etc.

Let us consider the first row. Here in order to reproduce the Standard Model we
restrict ourselves by the values of $Y$ equal to $0$ and $-2$. Next, the second
row must contain the only element with $Y = -1$. The third row contains two
elements with $Y = \frac{4}{3}$ and $Y = -\frac{2}{3}$. In the forth row we
have the only element with $Y = \frac{1}{3}$. This row completes the Standard
Model and we enter the rows related to its ultraviolet completion.

Before dealing with these next rows let us describe how parity conjugation of
spinors is incorporated in our model. We shall also remind what we call the
additional $Z_6$ symmetry in the Standard Model and how can it be continued to
the Hypercolor models.

\section{Parity conjugation}

Let us specify how parity conjugation $\cal P$ acts on the fermions. If only
two fermions $\chi^{\alpha}$ and $\eta_{\dot{\alpha}}$ are present, then ${\cal
P}\chi^{\alpha}(t,\bar{r}) = i \eta_{\dot{\alpha}}(t,-\bar{r}); {\cal P}
\eta_{\dot{\alpha}}(t,\bar{r})= i \chi^{\alpha}(t,-\bar{r})$. In our case we
require that for any configuration of $SU(N)$ ($N > 2$) indices there exist two
right - handed spinors and one $SU(2)$ doublet. The parity conjugation connects
each of the right handed spinors with a component of the $SU(2)$ doublet. Thus
\begin{eqnarray}
&& {\cal P}\Psi_0(t,\bar{r}) = i \Omega^1_{i_2}(t,-\bar{r})
\Theta^{i_2}_{-1}(t,-\bar{r}); {\cal P}\Psi_{-2} = i
\Omega^2_{i_2}\Theta^{i_2}_{-1};\nonumber\\
&& {\cal P}\Psi^{i_3}_{\frac{4}{3}} = i \Omega^1_{i_2} \Theta^{i_3
i_2}_{\frac{1}{3}}; {\cal P}\Psi^{i_3}_{-\frac{2}{3}} = i
\Omega^2_{i_2}\Theta^{i_3 i_2}_{\frac{1}{3}};\nonumber\\
&& {\cal P}\Psi^{i_4}_{Y_{4}} = i \Omega^1_{i_2} \Theta^{i_4 i_2}_{Y_{42}};
{\cal P}\Psi^{i_4}_{Y^\prime_{4}} = i
\Omega^2_{i_2}\Theta^{i_4 i_2}_{Y_{42}};\nonumber\\
&& {\cal P}\Psi^{i_4 i_3}_{Y_{43}} = i \Omega^1_{i_2} \Theta^{i_4
i_2}_{Y_{432}}; {\cal P}\Psi^{i_4 i_3}_{Y^\prime_{43}} = i
\Omega^2_{i_2}\Theta^{i_4 i_3 i_2}_{Y_{432}};\nonumber\\
&& ... \label{P}
\end{eqnarray}

Here $\Omega$  is an auxiliary $SU(2)$ field. $[\Omega^1]^*$ and $[\Omega^2]^*$
belong to the fundamental representation of $SU(2)$ subgroup of $G$.
 $U(1)$ subgroup of $G$ acts on $\Omega$ in such a way that $\Omega^1$ has hypercharge $1$ while $\Omega^2$
has hypercharge $-1$.

 Expression (\ref{P})
means that it is chosen dynamically, which component of $\Theta$ is connected
via parity conjugation with the given $\Psi$. The choice of parity conjugated
component of $\Theta$ is performed using an auxiliary field $\Omega$. The
physical sense of this field is that it peeks up the parity partner for each
right - handed spinor in a way that formally respects the chiral symmetry of
the theory.

\section{$\cal Z$ symmetry}

Here we follow the analysis of \cite{BVZ2003,Z2007,Z2007t}. Within the Standard
Model for any path $\cal C$, we may calculate the elementary parallel
transporters
\begin{eqnarray}
\Gamma &=& {\rm P} \, {\rm exp} (i\int_{\cal C} C^{\mu} dx^{\mu}) \nonumber\\
U &=& {\rm P} \, {\rm exp} (i\int_{\cal C} A^{\mu} dx^{\mu}) \nonumber\\
e^{i\theta} &=& {\rm exp} (i\int_{\cal C} B^{\mu} dx^{\mu}) ,\label{Sing}
\end{eqnarray}
where $C$, $A$, and $B$ are correspondingly $SU(3)$, $SU(2)$ and $U(1)$ gauge
fields of the Standard Model.

The parallel transporter correspondent to each fermion of the Standard Model is
the product of the elementary ones listed above. Therefore,  the elementary
parallel transporters are encountered in the theory only in the following
combinations: $e^{-2i\theta};\,U\, e^{-i\theta};\Gamma \, U \, e^{ \frac{i}{3}
\theta}; \Gamma \, e^{ -\frac{2i}{3} \theta}; \Gamma \, e^{ \frac{4i}{3}
\theta}$.

It can be easily seen \cite{BVZ2003} that {\it all} the listed combinations are
invariant under the following $Z_6$ transformations:
\begin{eqnarray}
 U & \rightarrow & U e^{i\pi N}, \nonumber\\
 \theta & \rightarrow & \theta +  \pi N, \nonumber\\
 \Gamma & \rightarrow & \Gamma e^{(2\pi i/3)N},
\label{symlat}
\end{eqnarray}
where $N$ is an arbitrary integer number.  This symmetry allows to define the
Standard Model with the gauge group $SU(3)\times SU(2) \times U(1)/{Z_6}$
instead of the usual $SU(3)\times SU(2) \times U(1)$.

It is worth mentioning that the additional discrete symmetry is rather
restrictive. Namely, for the Standard Model the requirement that the fermion
parallel transporters are invariant under $Z_6$ gives the condition for the
choice of the representations that are allowed for the Standard Model fermions.
Say, the left - handed $SU(2)$ doublets with zero hypercharge are forbidden.

The nature of the given additional  symmetry is related to the centers $Z_3$
and $Z_2$ of $SU(3)$ and $SU(2)$. This symmetry connects the centers of $SU(2)$
and $SU(3)$ subgroups of the gauge group. We suggest the following way to
continue this symmetry to the Hypercolor extension of the Standard Model.

We connect the center of the  Hypercolor group to the centers of $SU(3)$ and
$SU(2)$. Let $SU(K)$ be the Hypercolor group. Then the transformation
(\ref{symlat}) is generalized to \cite{Z2007t}
\begin{eqnarray}
 U & \rightarrow & U e^{i\pi N}, \nonumber\\
 \theta & \rightarrow & \theta +  \pi N, \nonumber\\
 \Gamma & \rightarrow & \Gamma e^{(2\pi i/3)N},\nonumber\\
 \Pi_4 & \rightarrow & \Pi_4 e^{(2\pi i/4)N},\nonumber\\
 \Pi_5 & \rightarrow & \Pi_5 e^{(2\pi i/5)N}, \nonumber\\
 \Pi_6 & \rightarrow & \Pi_6 e^{(2\pi i/6)N}, \nonumber\\
 ...
\label{symlatWS}
\end{eqnarray}
Here $\Pi_K$ is the $SU(K)$ parallel transporter. We construct our model in
such a way that the parallel transporters correspondent to the new fermions of
the theory are invariant under (\ref{symlatWS}). The resulting symmetry is
denoted by $\cal Z$ and enters expression (\ref{G}).

\section{Farhi - Susskind model}

Now let us consider the second four rows in (\ref{F}). We suggest them in the
form that represents $SU(4)$ Farhi - Susskind model of Technicolor \cite{FS}.
In this model the number of fermions is fixed, the $U(1)$ anomaly is absent but
the hypercharge assignment is not fixed. In order to make a choice  we apply
the continuation of the $Z_6$ symmetry found in the Standard Model.

We choose the hypercharge assignment here in such a way that:

1. Mass terms for the fermions proportional to $\Psi^+(t,\bar{r}) {\cal
P}\Psi(t,-\bar{r})$ are invariant under Electromagnetic $U(1)$. Therefore
\begin{eqnarray}
&& Y_4 = Y_{42} + 1; Y^\prime_5 = Y_{42} - 1;\nonumber\\
&& Y_{43} = Y_{432} + 1; Y^\prime_{43} = Y_{432} - 1;
\end{eqnarray}

2. Chiral anomaly should vanish. There are several types of the triangle
anomaly \cite{Weinberg}. Namely, in addition to the anomaly that may appear in
the Standard Model there may appear the anomaly of the following types:
\begin{eqnarray}
&&1) SU(4) - SU(4) - SU(4) \nonumber\\
&&2) SU(4) - SU(4) - U(1) \nonumber\\
&&3) SU(3) - SU(3) - U(1) \nonumber\\
&&4) SU(2) - SU(2) - U(1) \nonumber\\
&&5) U(1) - U(1) - U(1)
\end{eqnarray}

The anomaly of the first type is absent because the number of left - handed
fermions is equal to the number of the right - handed ones while both types of
fermions belong to the fundamental representation of $SU(4)$.

The anomaly of the second type is absent because
\begin{eqnarray}
&& 2 (Y_{42} + 3Y_{432}) - (Y_{42} + 1 + 3 (Y_{432} + 1)) - (Y_{42} -1 +
3(Y_{432}-1))= \nonumber\\
&&=  (Y_{42} + 3Y_{432})( 2 - 1 - 1) + 4  - 4 = 0 \label{SSU4}
\end{eqnarray}

The anomaly of the third type is absent because
\begin{eqnarray}
&& 6Y_{432} - (3 (Y_{432} + 1)) - (3(Y_{432}-1))= \nonumber\\
&&=  3Y_{432}( 2 - 1 - 1) + 3  - 3 = 0 \label{SSU3}
\end{eqnarray}

The anomaly of the fourth type is absent if
\begin{equation}
Y_{42} + 3Y_{432}= 0 \label{SSU2}
\end{equation}

The anomaly of the fifth type is absent if
\begin{eqnarray}
&& 2 (Y^3_{42} + 3Y^3_{432}) - ((Y_{42} + 1)^3 + 3 (Y_{432} + 1)^3) - ((Y_{42}
-1)^3 + 3(Y_{432}-1)^3) = \nonumber\\
&& =  2(Y^3_{42} + 3Y^3_{432})  - ((Y_{42} + 1 + Y_{42} - 1)^3 - 3(Y_{42} +
1)^2(Y_{42} - 1)  \nonumber\\
&&-  3(Y_{42} - 1)^2(Y_{42} + 1)) - 3((Y_{432} +1 + Y_{432}-1)^3 \nonumber\\
&&- 3(Y_{432} + 1)^2(Y_{432} - 1) - 3(Y_{432} - 1)^2(Y_{432} + 1)) \nonumber\\
&& = 2(Y^3_{42} + 3Y^3_{432})- (2Y_{42}^3 + 6Y_{42}) - 3 (2Y_{432}^3 +
6Y_{432}) \nonumber\\
&& = (Y^3_{42} + 3Y^3_{432})(2 - 2) - 6(Y_{42} + 3 Y_{432}) = 0\label{UUU}
\end{eqnarray}

This means that the sum of the hypercharge over the left - handed doublets
should be equal to zero:
\begin{equation}
Y_{42} + 3Y_{432} = 0
\end{equation}

3. The model should be invariant under the continuation of the $Z_{6}$ symmetry
of the Standard Model. Therefore, we come to the following equations:
\begin{eqnarray}
&&(\frac{2N}{4} + \frac{2N}{3} + N + Y_{432} N ) \, {\rm mod }\, 2 = 0  \nonumber\\
&&(\frac{2N}{4} + N + Y_{42} N) \, {\rm mod }\, 2 = 0  \nonumber\\
&&  Y_{42} + 3Y_{432} = 0
\end{eqnarray}

As a result the hypercharge assignment is the following \cite{Z2007t}. In the
$5$ - th row there are two elements with $Y_4 = \frac{1}{2}- 6K + 1$ and
$Y^\prime_4 = \frac{1}{2}- 6K -1$ (were $K$ is an arbitrary integer number). In
the $6$ -th row we have the only element with $Y_{42} = \frac{1}{2}- 6K$, where
$K$ is the same as in the previous row. In the $7$ - th row there are two
elements with $Y_{43} = -\frac{\frac{1}{2}- 6K}{3} + 1$ and $Y^\prime_{43} =
-\frac{\frac{1}{2}- 6K}{3} -1$. The $8$ -th row contains the only element with
$Y_{432} = -\frac{\frac{1}{2}- 6K}{3}$. Again, in these two rows $K$ is the
same as before.

For the definiteness let us list here the fermions for the choice $K = 0$.

\begin{eqnarray}
U(1): && \Psi_0, \Psi_{-2};\nonumber\\
U(1), SU(2): && \Theta^{i_2}_{-1}; \nonumber\\
U(1), SU(3): && \Psi^{i_3}_{\frac{4}{3}}; \Psi^{i_3}_{-\frac{2}{3}}; \nonumber\\
U(1), SU(2), SU(3): && \Theta^{i_3 i_2}_{\frac{1}{3}};\nonumber\\
U(1), SU(4): && \Psi^{i_4}_{\frac{3}{2}},\Psi^{i_4}_{-\frac{1}{2}};\nonumber\\
U(1), SU(2), SU(4): && \Theta^{i_4 i_2}_{\frac{1}{2}};\nonumber\\
U(1), SU(3), SU(4): && \Psi^{i_4 i_3}_{\frac{5}{6}}, \Psi^{i_4 i_3}_{-\frac{7}{6}};\nonumber\\
U(1), SU(2),SU(3),SU(4): &&  \Theta^{i_4 i_3 i_2}_{-\frac{1}{6}}; \nonumber\\
&&...\label{SMFS}
\end{eqnarray}

In the list (\ref{SMFS}) we have specified the Standard Model fermions and the
Farhi - Susskind model fermions. If the sequence (\ref{G}) is restricted by
these models only, the gauge group of the theory would be
\begin{equation}
SU(4) \otimes SU(3) \otimes SU(2) \otimes U(1)/ Z_{12} \label{Z12}
\end{equation}

The correspondence between our notations and the conventional ones is the
following (we consider the first generation only):
\begin{eqnarray}
 && \Psi_0 = \nu_R; \Psi_{-2} = e^-_R; {\cal P}\Psi_0(t,\bar{r}) = i \nu_L(t,-\bar{r}); {\cal P}\Psi_{-2} = i e^-_L ;\nonumber\\
 && \Psi^{i_3}_{\frac{4}{3}} = u_R; \Psi^{i_3}_{-\frac{2}{3}} = d_R; {\cal P}\Psi^{i_3}_{\frac{4}{3}} = i
 u_L;  {\cal P}\Psi^{i_3}_{-\frac{2}{3}} = i d_L;\nonumber\\
 && \Psi^{i_4}_{\frac{3}{2}} = N_R; \Psi^{i_4}_{-\frac{1}{2}}=E_R; {\cal P}\Psi^{i_4}_{\frac{3}{2}} = i
 N_L;  {\cal P}\Psi^{i_4}_{-\frac{1}{2}} = i E_L ;\nonumber\\
 && \Psi^{i_4 i_3}_{\frac{5}{6}} = U_R; \Psi^{i_4 i_3}_{-\frac{7}{6}}= D_R;
 {\cal P}\Psi^{i_4 i_3}_{\frac{5}{6}} = i U_L; {\cal P}\Psi^{i_4 i_3}_{-\frac{7}{6}} =
 i D_L.
\end{eqnarray}

 It is worth
mentioning that the fermions of the first generation listed here do not
diagonalize the mass matrix (see discussion of the fermion masses below).
Instead the certain linear combinations of the listed fermions diagonalize the
mass matrix thus giving rise to mixing angles and flavor changing amplitudes.

\section{Fermion masses}
In our construction we suppose that the formation of fermion masses is not
related to the chiral symmetry breaking due to the $SU(4)$ interactions. One
may suppose, for example, that the fermion masses appear at the energies much
higher than the energies at which the Hypercolor tower works. Let us suppose
that massless fermion is flying through a gas of objects such that inside them
the transition between the states related by parity conjugation may occur. In
particular, processes like that may happen within the objects, such that their
interior is organized in an unusual way. Namely, suppose, that inside that
objects the transformation that is seen from outside as a space inversion may
happen continuously. These objects, in turn, may have an origin of
gravitational nature. Probably, objects like that may belong to a class of
black holes supplemented by quantum effects.

Let the density of such objects be of the order of $\Lambda_h^3$ while their
size is about $M^{-1}_g$. Let the amplitude of the transition $\Psi \rightarrow
{\cal P}\Psi$ be proportional to the dimensionless constant $\beta$. Then it
can be easily  calculated that the massless fermion becomes massive with the
mass of the order of $m_{\Psi} \sim \beta \frac{\Lambda_h^3}{M_g^2}$. The
process like this happens in the Extended Technicolor theory, where massless
quark or lepton is flying through a gas of techniquarks. The ETC interactions
between them and the SM fermions occur at the distances $\sim
\frac{1}{M_{ETC}}$ while the density of technifermions that are condensed in
vacuum is of the order of $\Lambda_{TC}$. So, the SM fermion masses are
proportional to $\frac{\Lambda_{TC}^3}{M_{ETC}^2}$. However, in our
consideration we suppose that the given mechanism happens due to the physics at
the scales $\Lambda_h$ that may be extremely large. $\Lambda_h$ may even be of
the order of Plank mass. We do not require existence of the processes like ETC
transition between quarks and techniquarks. Therefore, our Hypercolor model
does not suffer from the problems specific for ETC models.

In order to incorporate the fermion masses to the theory we simply introduce
the mass term in the action in the following way. Let us denote the right -
handed fermions from the first column of (\ref{F}) as $U_A = U^{\alpha}_a =
\Psi^{A}_{Y_A+1}$, where $\alpha$ is the collection of indices of the subgroups
of (\ref{G}) while $a$ enumerates generations. The pair $(\alpha, a)$ that
identifies the fermion is denoted by  $A$. We denote the right - handed
fermions from the second column of (\ref{F}) as $D_A =
D^{\alpha}_a=\Psi^{A}_{Y_A-1}$. The left handed doublets are denoted $L_{A i} =
L^{\alpha}_{a i}= \Theta^{A i}_{Y_A}$. The hypercharge of the left - handed
fermion $A$ is denoted by $Y_A$. In order to provide invariance of the mass
term under the Electromagnetic $U(1)$ the correspondent right - handed fermions
have hypercharges $Y_A \pm 1$. The mass term is
\begin{eqnarray}
 && {\cal M} = i \sum_{U} M^U_{ab}  [U^{\alpha}_b(t, \bar{r})]^+ {\cal P}
U^{\alpha}_a(t, - \bar{r}) +i \sum_{D} M^D_{ab} [D^{\alpha}_b(t,
\bar{r})]^+{\cal P} D^{\alpha}_a(t, - \bar{r})  + c.c. \nonumber\\ &&=
 \sum_{U}
M^U_{ab} [U^{\alpha}_b]^+ \Omega^1_i L^{\alpha}_{a i}+ \sum_{D} M^D_{ab}
 [D^{\alpha}_b]^+\Omega^2_i L^{\alpha}_{a i} +
c.c. \label{MF1}
\end{eqnarray}

Here the sum is over the rows of (\ref{F}) to which $U$ and $D$ belong. The sum
over $a$, $b$, and $\alpha$ is also implied. Before diagonalization the mass
matrices $M^U$ and $M^D$ have block - diagonal forms. Each block corresponds to
a certain collection of Hypercolor subgroups of (\ref{G}) that act on the
correspondent fermion states. Both $M^U$ and $M^D$ can always be made diagonal
 (with real elements) via $U({\cal N})_L \otimes U({\cal N})_R$ rotations together with the
suitable redefinition of the $\theta$ - parameters in the $SU(N)$ theta terms
of the action: $M^U \rightarrow [{\cal K}_L^U]^+ [{M}^U]{\cal K}_R^U$; $M^D
\rightarrow [{\cal K}_L^D]^+ [{M}^D] {\cal K}_R^D$, where ${\cal K}_{L,R}^U \in
U({\cal N})$,  ${\cal K}_{L,R}^D \in U({\cal N})$. The dynamical part of the
fermion action is invariant under these transformations if ${\cal K}_L^U =
{\cal K}_L^D$. That's why in the charged weak currents the mixing matrix
$[{\cal K}_L^U]^+ {\cal K}_L^D$ appears that contains the usual CKM matrix of
the Standard Model.

Starting from the theory with diagonal real mass matrix using $U({\cal N})$
transformations we can always bring the theory to the form, in which mixing is
absent while the mass matrix is not diagonal but Hermitian. Namely, we can use
the transformation
\begin{equation}
M^U \rightarrow {\cal K}_L^U [{M}^U][{\cal K}_L^U]^+;\, M^D \rightarrow {\cal
K}_L^D [{M}^D] [{\cal K}_L^D]^+,\label{MU}
\end{equation}
where ${\cal K}_{L}^U $, and ${\cal K}_{L}^D$ are the same as above. Again
$M^U$ and $M^D$ have block - diagonal forms. The hermitian nature of the mass
matrix means, in particular, that the determinant of each mentioned block is
real. That's why (\ref{MU}) is not accompanied by the shift in  $\theta$ -
parameter of $SU(N)$ theta - term for any $N$. Below we always imply that
matrices $M^U$ and $M^D$ are Hermitian.

It is well - known that in such form of the theory the  $\theta$ parameter for
$SU(3)$ is negligible. The understanding of the reason why it is so is the
subject of the so - called strong CP problem of QCD. In our model we also imply
that the $\theta$ parameters for $SU(N)$ subgroups of (\ref{G}) vanish while
the mass matrix is Hermitian. In this language the analogue of the strong CP
problem of QCD is: Why is the mass matrix Hermitian while the theta - terms are
absent.

Actually, we do not try to give the answer to this question at the present
moment. We can only say, that if our supposition is correct and the mass matrix
is indeed Hermitian (while the theta - terms are absent), its hermitian nature
must originate from the energies at which (as described above) the transition
between the left - handed and the right - handed fermions occurs. Suppose that
the amplitude of such process for the transformation of the fermion $\psi_L$ to
the fermion $\phi_R$ is $A_{\psi_L \rightarrow \phi_R}$. On this language the
hermitian nature of the mass matrix means that the amplitude of the process
$\psi_R \rightarrow \phi_L$ is the same: $A_{\psi_R \rightarrow \phi_L} =
A_{\psi_L \rightarrow \phi_R}$. At the same time due to (\ref{MF1}) the
amplitude of the inverse process is given by the complex conjugate value:
$A_{\phi_R \rightarrow \psi_L} = A_{\phi_L \rightarrow \psi_R} = A^*_{\psi_L
\rightarrow \phi_R} = A^*_{\psi_R \rightarrow \phi_L}$.

The mass matrix as described by (\ref{MF1}) does not contain Majorana mass term
for the neutral right - handed neutrinos. In principle, this term can be added
to the model without causing any problems. It should have the form ${
M^{\nu_R}}_{ab} \epsilon_{\alpha \beta} \nu^{\alpha}_{R, a} \nu^{\beta}_{R,
b}$, where $a$ and $b$ indicate generations. Its high energy origin should
differ from that of the Dirac mass term. Namely, Majorana mass term is related
to the transformation of the right - handed neutrinos to their antiparticles.
(As described above the Dirac mass term is related to the transformation of the
fermion to its parity partner.) Further we do not focus on this question and
imply that the model can be updated if necessary in order to provide neutrino
masses in a proper way.

\section{The spontaneous breakdown of chiral symmetry}
 Let us suppose here that $SU(4)$ group is confining and gives rise to chiral
symmetry breaking. (Later we shall discuss the conditions under which this
happens.) Then the vacuum alignment \cite{Align} works in such a way that the
chiral condensates must be proportional to the only explicit $SU(2)$ variable
$\Omega$.

Let us define the field $\Phi$ as follows
\begin{equation}
\Phi_{i_2 B}^{1 A} = [\Theta^{ A i_2}_{Y_A}]^+ \Psi^{B}_{Y_B+1};\, \, \Phi_{i_2
B}^{2 A} = [\Theta^{A i_2}_{Y_{A}}]^+\Psi^{B }_{Y_{B} - 1}\label{phi0}
\end{equation}
where $A$, $B$ enumerate left handed fermions ($Y_A$, $Y_B$ are their
hypercharges). Right handed fermions $\Psi^{B}_{Y_B+1}$ belong to the first
column of (\ref{F}) while $\Psi^{B }_{Y_{B} - 1}$ belong to the second column.
In the previous section we defined index $A$ as a pair $(\alpha, a)$, where
$\alpha$ is the collection of $SU(K)$ indices ($SU(K)$ is a subgroup of
(\ref{G})) while $a$ enumerates generations. In this section the $SU(4)$
indices are ignored in this collection as we describe the effective theory,
which appears after Technicolor gauge field is integrated out. We imply the
mass matrix is diagonal in $SU(4)$ index. In (\ref{phi0}) summation over
$SU(4)$ index is implied. Below we omit indices $A$ and $B$ and imply that
$\Phi^i_j$ is ${\cal N} \times {\cal N}$ matrix for each $i$ and $j$.Then the
mass term in the action can be written as
\begin{equation}
{\cal M} = {\rm Tr} [\Phi^i_j]^+ {\cal M}^i_j + c.c.
\end{equation}
Here the mass matrix is ${\cal M}^i_j = [{\cal M}^i_j]^{A}_{B} = [{\cal
M}^i_j]^{a \alpha}_{b \alpha}$. It is expressed through $M^U_{ab}$ and
$M^D_{ab}$ as follows:
\begin{eqnarray}
 && [{\cal M}^1_i]^{a \alpha}_{b \beta} = M^U_{ab}\Omega^1_i \delta_{\alpha \beta}\nonumber\\
 && [{\cal M}^2_i]^{a \alpha}_{b \beta} = M^D_{ab}\Omega^2_i \delta_{\alpha \beta}
\end{eqnarray}

If all interactions but the Technicolor and the fermion masses are switched
off, then the Technicolor theory has the symmetry $SU(2{\cal N})_L\otimes
SU(2{\cal N})_R \otimes U(1)_V$, where $\cal N$ is the whole number of the left
handed doublets ($A,B = 1, ..., {\cal N}$). $U(1)_V$ acts identically on left -
handed and right handed fermions. ($U(1)_A$ is not a quantum symmetry due to
the anomaly.)  The effective action is

\begin{equation}
S(\Phi)  =  c_1 {\rm Tr} \, [{\cal D}\Phi]^+{\cal D}\Phi + V(\Phi)\label{V1}
\end{equation}
where the potential $V(\Phi)$ has the form

\begin{eqnarray}
 V(\Phi) & = &   c_2 ({\rm Tr}\,[\Phi^i_j]^+\Phi^i_j - \kappa^2)^2 + c_3 {\rm
Tr}\,[\Phi^{i_1}_{j_1}]^+\Phi^{i_1}_{j_2} [\Phi^{i_2}_{j_2}]^+\Phi^{i_2}_{j_1}
\nonumber\\&& -{\rm Tr}\,[\Phi^i_j]^+ {\cal M}^i_j - {\rm Tr}\,\Phi^i_j [{\cal
M}^i_j]^+ \label{EA1}
\end{eqnarray}

In the above expressions $\kappa$, and $c_k$ are unknown constants. The
derivative $\cal D$ contains all gauge fields but the Technicolor field. $\cal
M$ is the mass matrix. The terms with higher derivatives and higher powers of
$\Phi$ (for example, those that contain terms with the determinant) are not
relevant at low enough energies.

The physical processes described by (\ref{V1}) happen within the regions of
space with the typical size $\frac{1}{\Lambda_{TC}}$, where $\Lambda_{TC}$ is
the analogue of $\Lambda_{QCD}$. The duration of such processes can also be
estimated as  $\frac{1}{\Lambda_{TC}}$. Therefore the estimate of the mass term
fluctuation for the technifermion with mass $m_T$ is $\sim  m_T\,
\frac{1}{\Lambda^4_{TC}}\, \delta \Phi$. In case when the technifermion mass
$m_T$ is much larger than $\Lambda_{TC}$ this term dominates and the
fluctuations of the field $\Phi$ composed of this technifermion can be
estimated as $\delta \Phi \sim \frac{\Lambda^4_{TC}}{m_T}$. The energy
$\epsilon$ in the processes to be described by (\ref{V1}) must be much less
than the scale of $\delta \Phi$. This is needed in order to neglect nonlocal
terms with higher powers of the derivatives. That's why, (\ref{V1}) may be
relevant for $\frac{\epsilon}{\Lambda_{TC}} <<
[\frac{\Lambda_{TC}}{m_T}]^{1/3}$. In particular, for $m_T \sim 1000$ Tev,
$\Lambda_{TC} \sim 1$ Tev we would have $\epsilon << 100$ Gev while for $m_T
\sim 10$ Tev one should require $\epsilon << 300$ Gev. At the energies of the
order of the Electroweak scale $\sim 100$ Gev such extra massive technifermions
must not enter the effective action in the form (\ref{V1}). Actually, the
correspondent bilinear forms $\Phi$ defined in (\ref{phi0}) are suppressed by
the factors $\frac{m_T}{\Lambda_{TC}}$. That's why effective action (\ref{V1})
contains the field $\Phi$ composed of only those technifermions, the masses of
which is of the order of $1$ Tev and smaller. Therefore, fluctuations of $\Phi$
can be estimated as $\delta \Phi \sim \Lambda^3_{TC}$. Thus higher derivatives
in (\ref{V1})  are suppressed at the energies $\epsilon << \Lambda_{TC}$. Under
the same condition one can also neglect higher powers of $\Phi$  in
(\ref{EA1}). (The terms with the powers of $\Phi$ up to the fourth are needed
in order to provide nonzero vacuum average of $\Phi$.)

It is worth mentioning that (\ref{EA1}) is invariant under the chiral $U(1)_A$
if ${\cal M}=0$. This is relevant at $\epsilon << \Lambda_{TC}$ only. At
$\epsilon \sim \Lambda_{TC}$ the terms should appear that violate $U(1)_A$ (in
particular, the term that contains ${\rm det}\, \Phi$). The situation here
differs from that of the QCD with two or three light quarks. There the
effective potential contains the determinant of $\Phi$ already at the energies
much less than $\Lambda_{QCD}$ as the determinant contains the second and the
third powers of $\Phi$ respectively. In our case the expected number of
technifermions with masses around $1$ Tev is larger than $4$. That's why the
determinant contains higher powers of $\Phi$.

The first term in the effective action gives masses for $W$ and $Z$ bosons. The
next terms resolve the vacuum alignment problem. The true vacuum corresponds to
the minimum of the potential $V(\Phi)$. (We neglect here the perturbations due
to the Standard Model interactions and the $SU(K)$ Hypercolor interactions for
$K>4$.)

In order to demonstrate how the vacuum alignment works let us consider first
the simplified situation when ${\cal N} = 1$, ${\cal M}^j_i = m_j{\Omega}^j_i$
(values $m_{1}$ and $m_2$ are eigenvalues of $\cal M$; no sum over $j$ is
implied in the definition of $\cal M$).

Let the effective potential for the field $\Phi$ has the simplified form with
$c_3 = 0$:
\begin{equation}
V(\Phi)  =   c_2 ({\rm Tr} \, \Phi^+\Phi - \kappa^2)^2  - [\Phi^i_j]^* {\cal
M}^i_j - \Phi^i_j [{\cal M}^i_j]^*\label{EAs}
\end{equation}
 It is clear that the vacuum value of
$\Phi$ is proportional to ${\cal M}$. Thus
\begin{equation}
\Phi_{vac} = f {\cal M},\label{fM}
\end{equation}
where $f$ is the solution of the equation:
\begin{equation}
0 = 2 c_2 ( f^2 (m^2_1 + m_2^2) -  \kappa^2 ) f - 1
\end{equation}

In particular, if $\sqrt{m^2_1 + m_2^2} << 4c_2\kappa^3$, then $\Phi_{vac} =
(\frac{\kappa }{\sqrt{m^2_1 + m_2^2}} + \frac{1}{4 c_2 \kappa^2} +
O(\frac{\sqrt{m^2_1 + m_2^2}}{4c_2\kappa^3})) {\cal M}$. Rough estimates show,
however, that this limiting case is not too realistic. $\Lambda_{TC}$ is to be
of the order of $1$ Tev. This is due to the estimate of the weak gauge boson
masses: $M_Z, M_W \sim g_W \sqrt{c_2} \Phi_{vac} \sim 0.1
\frac{\Phi_{vac}}{\Lambda^2_{TC}}$. In case $\sqrt{m^2_1 + m_2^2} <<
4c_2\kappa^3 \sim \Lambda_{TC}$ we have $M_Z, M_W \sim 0.1 \Lambda_{TC}$. So,
in this case the technifermion masses are somewhere around $100$ Gev, or even
less. This situation cannot be considered as realistic. That's why we expect
that the masses of the technifermions are to be of the order of
$\Lambda_{TC}\sim 1$ Tev.

For any relation between technifermion masses and $\Lambda_{TC}$ due to
(\ref{fM}) we have
\begin{eqnarray}
\langle [\Theta^{i_4 i_2}_Y]^+ \Psi^{i_4}_{Y+1} \rangle \sim m_1 \Omega^1_{i_2} \nonumber\\
\langle [\Theta^{i_4 i_2}_{Y} ]^+ \Psi^{i_4}_{Y-1} \rangle \sim m_2
\Omega^2_{i_2}
\end{eqnarray}

Now we come back to the general case of ${\cal N} \ne 1$ and the effective
action (\ref{EA1}). Minimum of the effective potential is achieved at the
vacuum value $\Phi_{vac}$. Let us introduce the $SU(2 {\cal N})$ index ${\bf a}
= (A, i)$. Both ${\cal M}^{\bf a}_{\bf b}$ and $[\Phi_{vac}]^{\bf a}_{\bf b}$
are $2{\cal N}\times 2{\cal N}$ matrices. $\Phi_{vac}$ satisfies the equation
\begin{equation}
{\cal M} =  2 c_2 \Phi {\rm Tr}\,[\Phi]^+\Phi - 2 c_2 \kappa^2 \Phi + 2 c_3
\Phi [\Phi]^+\Phi \label{Eq}
\end{equation}

Using $SU(2 {\cal N})_{L,R}$ rotations we can always make the mass matrix
diagonal. Let us denote its diagonal elements $m_i$. It is easy to understand,
that  the matrix $\Phi_{vac}$ also becomes diagonal. We denote its diagonal
elements $\phi_i$. Thus (\ref{Eq}) leads to

\begin{equation}
m_i =  2 c_2 \phi_i \sum |\phi_i|^2 - 2 c_2 \kappa^2 \phi_i + 2 c_3 \phi_i
|\phi_i|^2
\end{equation}

In particular, in case $\frac{m_i (2{\cal
N}+\frac{c_3}{c_2})^{\frac{3}{2}}}{4\kappa^3 c_3} << 1$ we have
\begin{equation}
\phi_i  =  \frac{\kappa}{\sqrt{2 {\cal N} + \frac{c_3}{c_2}}} + \frac{1}{4
\kappa^2c_3} ((2{\cal N} + \frac{c_3}{c_2}) m_i - \sum m_i) + O([\frac{m_i
(2{\cal N}+\frac{c_3}{c_2})^{\frac{3}{2}}}{\kappa^3 c_3}]^2)
\end{equation}

(However, as for the simplified model considered above, this case is not
realistic and we actually have to consider technifermion masses to be of the
order of $1$ Tev.)

For any values of $m_i$ let us consider the vacuum value of $\Phi$ written in
the form $[({\Phi_{vac}})^i_j]^{a \alpha}_{b \beta}$, where $\alpha, \beta$
denote the collection of $...SU(6)\otimes SU(5)\otimes SU(3)$
 indices while $a,b$ enumerate generations. Symmetry properties of
$\Phi_{vac}$ are obvious. In Unitary gauge $\Omega = {\bf 1}$ the mass matrix
is such that ${\cal M}^2_1 = {\cal M}^1_2 = 0$. In this case
$(\Phi_{vac})^{1}_2 = (\Phi_{vac})^{2}_1 = 0$. One can easily see, that
$\Phi_{vac}$ preserves all symmetries of $\cal M$. Namely, let us rewrite the
mass matrix in the form $[{\cal M}^i_j]^{a \alpha}_{b \beta}$. Then $[{\cal
M}^i_j]^{a \alpha}_{b \beta}$  is nonzero only if $\alpha$ is identical to
$\beta$. Therefore,  $[({\Phi_{vac}})^i_j]^{a \alpha}_{b \beta} \ne 0$ only if
$\alpha$ coincides with  $\beta$. That's why Technicolor breaks the Electroweak
symmetry only.

In Unitary gauge the fields of $W$ and $Z$ bosons as well as the
Electromagnetic field $A$ are defined as usual. The mass matrix and
$[\Phi_{vac}]^j_j$ are invariant under the Electromagnetic $U(1)$ symmetry. At
the same time $[\Phi_{vac}]^j_j$ breaks Electroweak $SU(2)$ and the Hypercharge
$U(1)$. Therefore, the $W$ and $Z$ bosons acquire their masses while $A$
remains massless.

We suppose that coefficients $c_2$ and $c_3$ are real while matrix $\cal M$ is
Hermitian. That's why $[\Phi_{vac}]^i_j$ is Hermitian. We can define the four -
component spinors $u_{A}= \left(\begin{array}{c}\Psi^{A}_{Y_A + 1}\\\Theta^{A
1}_{Y_A}\end{array}\right)$ and $d_{A}= \left(\begin{array}{c}\Psi^{A}_{Y_A
-1}\\\Theta^{A 2}_{Y_A}\end{array}\right)$. Then the technipion condensate
vanishes:
\begin{eqnarray}
&&\langle \bar{u}_{A} \gamma_5 u_{B} \rangle = \langle [\Theta^{A 1}_{Y_A}]^+
\Psi^{A}_{Y_A + 1} - [\Psi^{A}_{Y_A+1}]^+\Theta^{A 1}_{Y_A} \rangle = 0
\nonumber\\ && \langle \bar{d}_{A}\gamma_5  d_{B} \rangle = \langle [\Theta^{A
2}_{Y_A}]^+ \Psi^{A}_{Y_A - 1} - [\Psi^{A}_{Y_A-1}]^+\Theta^{A 2}_{Y_A} \rangle
= 0 \label{TP}
\end{eqnarray}
The physical sense of (\ref{TP}) is trivial. It means that the Technicolor
vacuum is invariant under the space inversion.

The next step in the consideration of the vacuum alignment would be to take
into account small perturbations due to the Standard Model interactions (and
due to the other interactions corresponding to the subgroups of (\ref{G})). It
was found in \cite{Align} that due to the Standard Model interactions the
conventional form of the chiral condensate appears. We suppose, that the higher
subgroups of (\ref{G}) do not introduce anything new. Up to this assumption we
come to the conclusion that in our case Technicolor breaks the Electroweak
symmetry properly.

\section{The further continuation}

The next step of our investigation is the analysis of the sequence
(\ref{SMFS}). Let us notice that the second two rows are actually the copy of
the first two rows supplemented by an additional $SU(3)$ index. Next, the
second four rows are again the copy of the first four rows supplemented by an
additional $SU(4)$ index. Let us suppose that this process is repeated
infinitely. Then, the next $8$ rows in the sequence are added in the form:

\begin{eqnarray}
... \nonumber\\
U(1), SU(5): && \Psi^{i_5}_{Y_5}, \Psi^{i_5}_{Y_5^\prime};\nonumber\\
U(1), SU(2), SU(5): && \Theta^{i_5 i_2}_{Y_{52}}; \nonumber\\
U(1), SU(3), SU(5): && \Psi^{i_5 i_3}_{Y_{53}}; \Psi^{i_5 i_3}_{Y_{53}^{\prime}}; \nonumber\\
U(1), SU(2), SU(3), SU(5): && \Theta^{i_5 i_3 i_2}_{Y_{532}};\nonumber\\
U(1), SU(4), SU(5): && \Psi^{i_5 i_4}_{Y_{54}},\Psi^{i_5 i_4}_{Y_{54}^{\prime}};\nonumber\\
U(1), SU(2), SU(4), SU(5): && \Theta^{i_5 i_4 i_2}_{Y_{542}};\nonumber\\
U(1), SU(3), SU(4), SU(5): && \Psi^{i_5 i_4 i_3}_{Y_{543}}, \Psi^{i_5 i_4 i_3}_{Y_{543}^\prime};\nonumber\\
U(1), SU(2),SU(3),SU(4), SU(5): &&  \Theta^{i_5 i_4 i_3 i_2}_{Y_{5432}}; \nonumber\\
&&...\label{SMFS8}
\end{eqnarray}

Again, we choose the hypercharge assignment in such a way that:

1. Mass terms for the fermions proportional to $\Psi^+(t,\bar{r}) {\cal
P}\Psi(t,-\bar{r})$ are invariant under Electromagnetic $U(1)$. Therefore
\begin{eqnarray}
&& Y_5 = Y_{52} + 1; Y^\prime_5 = Y_{52} - 1;\nonumber\\
&& Y_{53} = Y_{532} + 1; Y^\prime_{53} = Y_{532} - 1;\nonumber\\
&& Y_{54} = Y_{542} + 1; Y^\prime_{54} = Y_{542} - 1;\nonumber\\
&& Y_{543} = Y_{5432} + 1; Y^\prime_{543} = Y_{5432} - 1\label{RL}
\end{eqnarray}

2. Chiral anomaly should vanish. The consideration is similar to that of
section $5$. Namely, there may appear the new anomalies of the following types:
\begin{eqnarray}
&&1) SU(5) - SU(5) - SU(5) \nonumber\\
&&2) SU(5) - SU(5) - U(1) \nonumber\\
&&3) SU(4) - SU(4) - U(1) \nonumber\\
&&4) SU(3) - SU(3) - U(1) \nonumber\\
&&5) SU(2) - SU(2) - U(1) \nonumber\\
&&6) U(1) - U(1) - U(1)
\end{eqnarray}

The anomaly of the first type vanishes because the number of left - handed
fermions is equal to the number of the right - handed ones while both types of
fermions belong to the fundamental representation of $SU(5)$.

The anomalies of the second, the third, and the fourth types vanish due to
(\ref{RL}).

The anomalies of the fifth and the sixth types vanish if the sum of the
hypercharge over left - handed doublets is zero. Thus
\begin{equation}
Y_{52} + 3Y_{532} + 4Y_{542} + 4 \times 3 \times Y_{5432} = 0
\end{equation}

3. The model must be invariant under the further continuation of the $Z_{12}$
symmetry of (\ref{Z12}). Therefore
\begin{eqnarray}
&&(\frac{2N}{5} + \frac{2N}{4} + \frac{2N}{3} + N + Y_{5432} N ) \, {\rm mod }\, 2 = 0  \nonumber\\
&&(\frac{2N}{5} + \frac{2N}{4} + N + Y_{542} N) \, {\rm mod }\, 2 = 0  \nonumber\\
&&(\frac{2N}{5} + \frac{2N}{3} + N + Y_{532} N)\, {\rm mod }\, 2 = 0  \nonumber\\
&&(\frac{2N}{5} +  N + Y_{52}N) \, {\rm mod }\, 2 = 0  \nonumber\\
&&  Y_{52} + 3Y_{532} + 4Y_{542} + 4 \times 3 \times Y_{5432} = 0
\end{eqnarray}

The solution is
\begin{eqnarray}
&& Y_{52} = \frac{3}{5} - 2(3 K_{532} + 4 K_{542} + 12 K_{5432}); Y_5 = Y_{52}
+ 1; Y^\prime_5 =
Y_{52} - 1;\nonumber\\
&& Y_{532} = \frac{29}{15} + 2 K_{532}; Y_{53} = \frac{44}{15}+ 2 K_{532}; Y^\prime_{53} =  \frac{14}{15}+ 2 K_{532};\nonumber\\
&& Y_{542} = \frac{1}{10}+ 2 K_{542}; Y_{54} = \frac{11}{10}+ 2 K_{542}; Y^\prime_{54} = -\frac{9}{10}+ 2 K_{542};\nonumber\\
&& Y_{5432} = -\frac{17}{30}+ 2 K_{5432}; Y_{543} =  \frac{13}{30}+ 2 K_{5432};
Y^\prime_{543} = -\frac{47}{30}+ 2 K_{5432}
\end{eqnarray}

Here $K_{532}, K_{542}, K_{532}$ are arbitrary integer numbers.

\section{Higher Hypercolor groups}

In this section we derive the hypercharge assignment for all fermions of our
model. We require that the chiral anomaly is absent and the additional $\cal Z$
symmetry is preserved. Actually, the fact that there exists such a solution is
 nontrivial. A priory it is not clear that it is possible to satisfy both
 requirements simultaneously.

Let us continue the sequence (\ref{SMFS8}) infinitely. It has the form:
\begin{eqnarray}
... \nonumber\\
U(1), SU(5): && \Psi^{i_5}_{Y_5}, \Psi^{i_5}_{Y_5^\prime};\nonumber\\
U(1), SU(2), SU(5): && \Theta^{i_5 i_2}_{Y_{52}}; \nonumber\\
U(1), SU(3), SU(5): && \Psi^{i_5 i_3}_{Y_{53}}; \Psi^{i_5 i_3}_{Y_{53}^{\prime}}; \nonumber\\
U(1), SU(2), SU(3), SU(5): && \Theta^{i_5 i_3 i_2}_{Y_{532}};\nonumber\\
U(1), SU(4), SU(5): && \Psi^{i_5 i_4}_{Y_{54}},\Psi^{i_5 i_4}_{Y_{54}^{\prime}};\nonumber\\
U(1), SU(2), SU(4), SU(5): && \Theta^{i_5 i_4 i_2}_{Y_{542}};\nonumber\\
U(1), SU(3), SU(4), SU(5): && \Psi^{i_5 i_4 i_3}_{Y_{543}}, \Psi^{i_5 i_4 i_3}_{Y_{543}^\prime};\nonumber\\
U(1), SU(2),SU(3),SU(4), SU(5): &&  \Theta^{i_5 i_4 i_3 i_2}_{Y_{5432}}; \nonumber\\
&&... \nonumber\\
U(1), ... , SU(K): && \Psi^{i_K ... }_{Y_{K...}}, \Psi^{i_K ...}_{Y_{K ...}^\prime};\nonumber\\
U(1), SU(2), ... , SU(K): &&  \Theta^{i_K ... i_2}_{Y_{K...2}}; \nonumber\\
... \label{SMFS100}
\end{eqnarray}

Now we require that the chiral anomaly is absent while the gauge group is
(\ref{G}), where $\cal Z$ is defined by (\ref{symlatWS}). Below we prove that
{\bf the necessary hypercharge assignment is}
\begin{eqnarray}
&& Y_2  =  -1 \nonumber\\
&&Y_{i_1 i_2 i_3 ... i_{M-1} i_M 2} = -1 + 2(1 - \frac{1}{i_M})  +  2 \sum_{k =
1}^{M-1}[\theta(i_k - i_{k+1} - 1) - \frac{1}{i_k}] + 2 N_{i_1 i_2 i_3 ...
i_{M-1} i_M 2}
\,\nonumber\\
&&Y_{ i j ... l} = Y_{ i j ... l 2} + 1; \, Y^\prime_{ i j ... l }  = Y_{ i j
... l 2} - 1\label{Y0}
\end{eqnarray}
where $\theta(x) = 1 \, {\rm for} \,x>0; \, \theta(x) = 0 \, {\rm for} \,x\le
0$. In the second row $\, M \ge 1$. For any $K$ integer numbers $N_{i_1 i_2 i_3
... i_{M-1} i_M 2}$ entering (\ref{Y0}) must satisfy the equation
\begin{equation}
\sum_{K > i > j > ... > l > 2} i j ... l \, N_{K ij...l2} = 0 \label{Y11}
\end{equation}
Here the sum is over any (unordered) sets of different integer numbers
$i,j,...,l$ such that $2<i,j,...,l <K$.

{\bf The proof} is as follows. First of all, if (\ref{symlatWS}) is the
symmetry of the theory then the recursion relations take place:
\begin{equation}
Y_{K i j ... l 2} = Y_{i j ... l 2} - \frac{2}{K} + 2 M_{K ij...l2}; Y_{K i j
... l} = Y_{K i j ... l 2} + 1; Y^\prime_{K i j ... l }  = Y_{K i j ... l 2} -
1,
\end{equation}
where $M_{K ij...l2}$ is an integer number.

Let us require that for any $K$
\begin{equation}
\sum_{K > i > j > ... > l > 2} i j ... l \, Y_{K i j ... l 2} = 0,\label{Y}
\end{equation}
 This means that the chiral anomaly is absent  even if the
sequence (\ref{G}) is ended at the $SU(K)$ factor with any value of $K$.

Namely. there may appear the new anomalies of the following types:
\begin{eqnarray}
&&1) SU(N) - SU(N) - SU(N),\, N > 2 \nonumber\\
&&2) SU(N) - SU(N) - U(1),\, N>2 \nonumber\\
&&3) SU(2) - SU(2) - U(1) \nonumber\\
&&4) U(1) - U(1) - U(1)
\end{eqnarray}

The anomaly of the first type vanishes because the number of left - handed
fermions is equal to the number of the right - handed ones while both types of
fermions belong to the fundamental representation of $SU(N)$.

The anomalies of the second type vanish because $Y_{ i j ... l} = Y_{ i j ... l
2} + 1; \, Y^\prime_{ i j ... l }  = Y_{ i j ... l 2} - 1$.

The anomalies of the third and the fourth types vanish if the sum of the
hypercharge over left - handed doublets is zero. This leads to (\ref{Y}).

Below we prove that for any $K$ integer numbers $M_{K ij...l2}$ can be chosen
in such a way, that (\ref{Y}) is satisfied. Let $\sum_{K^\prime > i > j > ... >
l > 2} i j ... l Y_{K^\prime i j ... l 2} = 0$ for $K^\prime < K$ (this was
demonstrated already for $K^\prime = 4$.). Then
\begin{eqnarray}
&& \sum_{K > i > j > ... > l > 2} i j ... l \, Y_{K i j ... l 2} = \sum_{K > i
> j > ... > l > 2} i j ... l \, Y_{i j ... l 2} \nonumber\\ && - \frac{2}{K}
\sum_{K > i > j > ... > l > 2} i j ... l + 2\sum_{K > i > j > ... > l > 2} i j ... l \, M_{K ij...l2}\nonumber\\
&& = - \frac{2}{K} \sum_{K > i > j > ... > l > 2} i j ... l  + 2\sum_{K> i> j >
... > l > 2} i j ... l \, M_{K ij...l2}\nonumber\\ && = - 2 \frac{K!}{3!K} +
2\sum_{K > i > j
> ... > l > 2} i j ... l \, M_{K ij...l2} \label{KM}
\end{eqnarray}

Here we used the identity
\begin{equation}
\sum_{K > i > j > ... > l > 2} i j ... l = \frac{K!}{3!} \label{K}
\end{equation}

The derivation of (\ref{K}) is as follows.  Suppose that (\ref{K}) is valid for
a certain number of $K$ (this is evident, for example, for $K = 4$). Then
\begin{eqnarray}
 \sum_{2<i j ... l<K+1} i j ... l && = 1 + 3\frac{3!}{3!} +  4\frac{ 4!}{3!} +
5\frac{ 5!}{3!} + ... +  K\frac{K!}{3!}\nonumber\\ && = \frac{1}{3!} ( 3! + 3!
3 + 4 !4 + ... + K !K)\nonumber\\ && = \frac{1}{3!} ( 4! + 4! 4 + ... + K! K)
\nonumber\\ && = \frac{1}{3!}  (K+1)!
\end{eqnarray}
From (\ref{KM}) it is clear that for $K>3$ it is always possible to choose one
of the values $M_{K ij...l2}$ in such a way that (\ref{Y}) is satisfied. The
same statement for $K = 3$ is also valid as follows from the consideration of
the Standard Model.

Let us know introduce the following notations:
\begin{eqnarray}
M_{K ij...l2} = M_{K ij...l2}^\prime + 1,\, {\rm for}\, \, K-1 > i > j > ... >
l > 2;\nonumber\\
M_{K ij...l2} = M_{K ij...l2}^\prime,\, {\rm for}\, \, K-1 = i > j > ... > l >
2
\end{eqnarray}

Then
\begin{equation}
-\frac{K!}{3!K} + \sum_{K > i > j > ... > l > 2} i j ... l \, M_{K ij...l2}  =
\sum_{K > i > j > ... > l > 2} i j ... l \, M^\prime_{K ij...l2}
\end{equation}

The relations that define the fermion hypercharges can be rewritten in the
following way:
\begin{eqnarray}
&&Y_{K i j ... l} = Y_{K i j ... l 2} + 1; \, Y^\prime_{K i j ... l }  = Y_{K i
j
... l 2} - 1,\nonumber\\
&&Y_{K i j ... l 2} = Y_{i j ... l 2} - \frac{2}{K} + 2 + 2 M^\prime_{K
ij...l2} \,\nonumber\\ && ( {\rm for}\, \, K-1
> i > j > ... > l > 2, \, {\rm or} \, K = 3);\nonumber\\
&&Y_{K i j ... l 2} = Y_{i j ... l 2} - \frac{2}{K}  + 2 M^\prime_{K ij...l2}
\, \nonumber\\&&({\rm for}\, \, K-1 = i > j > ... > l > 2)
\end{eqnarray}
Here integer numbers $M^\prime_{K ij...l2}$ are chosen in such a way that
\begin{equation}
\sum_{K > i > j > ... > l > 2} i j ... l \, M^\prime_{K ij...l2} = 0 \label{Y1}
\end{equation}

Finally we come to the solution of (\ref{Y}) in the form (\ref{Y0}). In
particular, the choice $N_{i_1 i_2 i_3 ... i_{M-1} i_M 2}=0$ corresponds to
\begin{equation}
Y_{i_1 i_2 i_3 ... i_{M-1} i_M 2} = -1 + 2(1 - \frac{1}{i_M}) +  2 \sum_{k =
1}^{M-1}[\theta(i_k - i_{k+1} - 1) - \frac{1}{i_k}]
\end{equation}

 Thus the additional symmetry (\ref{symlatWS}) fixes the hypercharge
assignment up to the choice of integer numbers $N_{i_1 i_2 i_3 ... i_{M-1} i_M
2}$ such that (\ref{Y11})  is satisfied. We cannot eliminate this uncertainty
at this stage.

\section{The relation between the discrete $\cal Z$ symmetry and the monopole content of
the Unified model.}

In the previous sections we apply the additional $\cal Z$ symmetry to the
construction of the Hypercolor model. The main reason for us to do so is that
the $Z_6$ symmetry of the Standard Model seems to us so peculiar, that we
expect it must be present in a certain form in the completion of the Standard
Model. Of course, the form (\ref{symlatWS}) of the continuation of this
symmetry is just our supposition.

The observability of the additional $\cal Z$ symmetry of the Hypercolor model
must be related to the topological objects existing within the more fundamental
theory that has our tower of Hypercolor groups as a description of the low
energy approximation. Let us consider the construction of the monopole
configuration (see, for example, \cite{Z2007}) of the hypothetical Unified
model.

We fix the closed surface $\Sigma$ in $4$-dimensional space $R^4$. For any
closed loop $\cal C$, which winds around this surface, we may calculate the
Wilson loops $\Pi_K = {\rm P} \,  {\rm exp} (i\int_{\cal C} H_K^{\mu}
dx^{\mu})$, $\Gamma = {\rm P} \,  {\rm exp} (i\int_{\cal C} C^{\mu} dx^{\mu})$,
$U = {\rm P} \, {\rm exp} (i\int_{\cal C} A^{\mu} dx^{\mu})$, and $e^{i\theta}
= {\rm exp} (i\int_{\cal C} B^{\mu} dx^{\mu})$, where $H_K$, $C$, $A$, and $B$
are correspondingly $SU(K)$, $SU(3)$, $SU(2)$ and $U(1)$ gauge fields of the
model. In the usual realization of the Hypercolor model with the gauge group
$... \otimes SU(3)\otimes SU(2) \otimes U(1)$ such Wilson loops should tend to
unity, when the length of $\cal C$ tends to zero ($|{\cal C}| \rightarrow 0$).
However, in the $... \otimes SU(3)\otimes SU(2) \otimes U(1)/{\cal Z}$ gauge
theory the following values of the Wilson loops are allowed at $|{\cal C}|
\rightarrow 0$:
\begin{eqnarray}
\Pi_K &=&  e^{N \frac{2\pi i}{K}}\nonumber\\ \Gamma &=&  e^{N \frac{2\pi
i}{3}}\nonumber\\
U &=& e^{N \pi i}\nonumber\\
e^{i\theta} &=& e^{N \pi i},\label{Sing}
\end{eqnarray}
where $N \in Z$. Then the surface $\Sigma$ may carry $SU(K)/Z_K$ flux $\pi [N\,
{\rm mod}\,K]$.

Any configuration with the singularity of the type (\ref{Sing}) could be
eliminated via a singular gauge transformation. Therefore the appearance of
such configurations in the theory with the gauge group $... \otimes
SU(3)\otimes SU(2) \otimes U(1)/{\cal Z}$ does not influence the dynamics.

Now let us consider an open surface $\Sigma$. Let the small vicinity of its
boundary $U(\partial \Sigma)$ represent a point - like soliton state of the
unified theory. This means that the fields of the Hypercolor model are defined
now everywhere except $U(\partial \Sigma)$. Let us consider such a
configuration, that for infinitely small contours $\cal C$ (winding around
$\Sigma$) the mentioned above Wilson loops are expressed as in (\ref{Sing}).
For $N \ne 0$ it is not possible to expand the definition of such a
configuration to $U(\partial \Sigma)$. However, this could become possible
within the unified model if the gauge group of the Hypercolor model $...\otimes
SU(3)\otimes SU(2) \otimes U(1)/{\cal Z}$ is embedded into the simply connected
group $\cal H$. This follows immediately from the fact that any closed loop in
such $\cal H$ can be deformed smoothly to a point and this point could be moved
to unity. Actually, for such $\cal H$ we have $\pi_2({\cal H}/[... \otimes
SU(3)\otimes SU(2) \otimes U(1)/{\cal Z}]) = \pi_1(... \otimes SU(3)\otimes
SU(2) \otimes U(1)/{\cal Z})$. This means that in such unified model the
monopole-like soliton states are allowed. The configurations with (\ref{Sing})
and $N\ne 0$ represent fundamental monopoles of the unified model\footnote{The
configurations of this kind were considered, for example, in \cite{Z6}, where
they represent fundamental monopoles of the $SU(5)$ unified model.}. The other
monopoles could be constructed of the fundamental monopoles as of the building
blocks.  In the unified model, which breaks down to the Hypercolor model with
the gauge group $... \otimes SU(3)\otimes SU(2) \otimes U(1)$ such
configurations for $N\ne 0$ are simply not allowed.

The unified model, which breaks down to the Hypercolor model  with the gauge
group $... \otimes SU(3)\otimes SU(2) \otimes U(1)$ also contains monopoles
because $\pi_2({\cal H}/[... \otimes SU(3)\otimes SU(2) \otimes U(1)]) =
\pi_1(... \otimes SU(3)\otimes SU(2) \otimes U(1)) = Z$. They correspond to the
Dirac strings with $ \int_{\cal C} B^{\mu} dx^{\mu} = 2 Q_{max} \pi K, K\in Z$.
(We  suppose here that the hypercharges of the fermions are rational numbers
$\frac{P}{Q}$ with integer $P$ and $Q$, and the maximal value of $Q$ is
$Q_{max}$.) Those monopoles should be distinguished from the monopoles  of the
Hypercolor model with the additional discrete symmetry via counting their
hypercharge $U(1)$ magnetic flux.

Using an analogy with t'Hooft - Polyakov monopoles\cite{HooftPolyakov} we can
estimate masses of  the monopole of the hypothetical Unified theory (in the
presence of $\cal Z$ symmetry)  as

\begin{equation}
M_N \sim 4 \pi \Lambda_h \, [\frac{1}{g_{U(1)}} + \sum_{K} \frac{1}{g_{SU(K)}}]
\,N \label{MN}
\end{equation}
In this sum the term corresponding to $SU(K)$ is absent if $N/K\in Z$ because
in this case the monopole does not carry $SU(K)/Z_K$ flux. Here $\Lambda_h$ is
the Unification scale and $ \pi N$ is the Hypercolor flux carried by the
monopole.

It is worth mentioning that the usual magnetic flux of the given monopoles is
$2 \pi$. This follows simply from the expression for the Electromagnetic field
through the $SU(2)$ field $A$ and the hypercharge field $B$: \begin{equation}
A_{\rm em} = 2 B + 2 \,{\rm sin}^2\, \theta_W (A_3 - B) \label{flux}
\end{equation}
The mentioned above monopoles have nontrivial $SU(2)/Z_2$ flux that cancels the
hypercharge flux within the second term of (\ref{flux}). That's why their usual
flux (with respect to the Electromagnetic $U(1)$) corresponds only to the first
term in (\ref{flux}) and is equal to $2 \pi$.

If hypercharge flux is proportional to $2 \pi$ then the monopole must not
necessarily carry $SU(2)/Z_2$ flux. In this case the field $A_3$ does not give
any contribution to the Electromagnetic flux. And the monopole may carry the
usual magnetic flux proportional to $ 4 \pi \,{\rm cos}^2 \theta_W $ due to
both terms of (\ref{flux}).

Let us suppose that at the Unification scale all couplings become close to each
other: $g^2 = g_{U(1)}^2(\Lambda_h) \sim g_{SU(2)}^2(\Lambda_h)\sim ... \sim
g_{SU(K)}^2(\Lambda_h)\sim ...  $. (In general case this is not necessarily so.
For example, in the models of the so - called Petite Unification that occurs at
a Tev scale the gauge couplings are not close to each other \cite{PUT}.) We
also suppose that the sequence  (\ref{G}) is ended at the Hypercolor group
$SU(K_{max})$. Then
\begin{equation}
M_N > \frac{4 \pi \Lambda_h}{g} K_{max} \label{MN2}
\end{equation}

So, we come to the conclusion that in case the $\cal Z$ symmetry is present the
appearance of monopoles in a hypothetical Unified theory is highly suppressed.

Let us now consider the case, when the gauge group of the Hypercolor tower is
$G = ... \otimes SU(5)\otimes SU(4) \otimes SU(3) \otimes SU(2) \otimes U(1)$
while the sequence of fermions is still given by (\ref{SMFS100}). The
hypercharge assignment is such that any of the subgroups of $\cal Z$ is not a
symmetry of the theory. We again suppose that the hypercharges of the fermions
are rational numbers $\frac{P}{Q}$ with integer $P$ and $Q$, and the maximal
value of $Q$ is $Q_{max}$. The hypercharge of the left - handed quarks is
$\frac{1}{3}$. That's why $Q_{max} \ge 3$. The smallest possible hypercharge
flux of the monopole is $2 Q_{max} \pi$. The groups $SU(N)$ are not involved in
such monopole configurations. The magnetic flux of the monopole is proportional
to $ 4 Q_{max} \pi \,{\rm cos}^2 \theta_W $. The estimate of the minimal
monopole mass is
\begin{equation}
M \sim \frac{8 Q_{max} \pi \Lambda_h}{g}\label{Mm}
\end{equation}

In the case, when a certain subgroup of $\cal Z$ serves as a symmetry group
given by the hypercharge assignment of (\ref{SMFS100}), the
 minimal monopole mass may be larger than (\ref{Mm}). And if this subgroup contains $Z_2$ then the
 monopoles may appear that carry usual magnetic flux proportional to $Z_2$. In order to increase
 the minimal monopole mass one may increase $Q_{max}$ or to make larger and larger subgroup of $\cal Z$ the symmetry of the
 theory. That's why the requirement that the whole $\cal Z$ is the symmetry is
 one of the ways to suppress monopoles (although, not the only one).

For the definiteness, let us demonstrate how, in principle, the Technicolor and
the Standard Model interactions may be unified in a common gauge group. Here we
do not consider higher Hypercolor groups and follow the construction suggested
in \cite{Z2007t}. We do not discuss  the details of the breakdown mechanism and
how the chiral anomaly cancellation is provided within the given scheme of
Unification. Our aim here is to demonstrate how the additional discrete
symmetry (\ref{symlatWS}) may appear during the breakdown of Unified gauge
symmetry.

Let $SU(10)$ be the Unified gauge group. The breakdown pattern is
$SU(10)\rightarrow SU(4)\otimes SU(3)\times SU(2) \times U(1)/Z_{12}$. We
suppose that at low energies the $SU(10)$ parallel transporter has the form:

\begin{equation}
\Omega = \left( \begin{array}{c c c c}
\Theta e^{-\frac{2i\theta}{4}} & 0 & 0 & 0 \\
0 & \Gamma^+ e^{\frac{2i\theta}{3}} & 0 & 0 \\
0 & 0 & Ue^{-i\theta} & 0\\
0 & 0 & 0 & e^{2i\theta}
\end{array}\right)\in SU(10), \label{U(10)}
\end{equation}

The fermions of each generation $\Psi^{i_1 ... i_N}_{j_1 ... j_K}$  carry
indices $i_k$ of the fundamental representation of $SU(10)$ and the indices
$j_k$ of the conjugate representation. They may be identified with the Standard
Model fermions and Farhi - Susskind fermions as follows (we consider here the
first generation only):

\begin{eqnarray}
&& \Psi^{10} = e^c_R; \, \Psi_{10}^{10} = \nu_R; \, \Psi^{i_2} =
\left(\begin{array}{c} \nu_L
\\ e^-_L\end{array}\right) ;\nonumber\\
&&\Psi^{i_3} = d^c_{i_3,R} ; \, \Psi^{i_3}_{10} = u^c_{i_3,R} ;\, \Psi^{i_2}_{
i_3} =
\left(\begin{array}{c} u^{i_3}_L \\ d^{i_3}_L \end{array}\right) ;\nonumber\\
&&\Psi_{i_4} = E^c_{i_4,R} ; \, \Psi_{10, i_4} = N^c_{i_4,R} ;\, \Psi^{i_2 i_4}
=
\left(\begin{array}{c} N^{i_4}_L \\ E^{i_4}_L \end{array}\right) ;\nonumber\\
&&\Psi^{i_3}_{i_4} = D^c_{i_3 i_4,R} ; \, \Psi_{10, i_4}^{i_3} = U^c_{i_3
i_4,R} ;\, \Psi^{i_2 i_4}_{i_3} = \left(\begin{array}{c} U^{i_3 i_4}_L \\
D^{i_3 i_4}_L
\end{array}\right) \nonumber\\&& (i_2 = 8,9; \, i_3 =
5,6,7;\, i_4 = 1,2,3,4);
\end{eqnarray}

Here the charge conjugation is defined as follows: $f^c_{\dot{\alpha}} =
\epsilon_{\alpha \beta} [f^{\beta}]^*$.

In principle the fermion content of the Unified model should be chosen in such
a way that the anomalies are cancelled. Moreover, some physics should be added
in order to provide "unnecessary" fermions with the masses well above $1$ Tev
scale. Besides, one must construct the unambiguous theory in such a way that at
low energies the parallel transporters indeed have the form (\ref{U(10)}). Let
us suppose that this program is fulfilled. Then all parallel transporters in
the theory are invariant under (\ref{symlatWS}) in a natural way. The gauge
group $SU(10)$ is simply connected. That's why the Unified theory should
contain monopole - like topological objects. As it was already mentioned, their
masses and magnetic fluxes are related essentially to the $\cal Z$ symmetry.

\section{Dynamics}

Now let us consider the dynamics of Technicolor. It is related in a usual way
to the number of fermions $N_f$. Namely, the beta - function in one loop
approximation has the form:
\begin{equation}
\beta_{SU(K)}(\alpha) = - \frac{11 K - 2 N_f}{6 \pi} \alpha^2,\label{BETA}
\end{equation}
where $\alpha = \frac{g^2_{SU(K)}}{4\pi}$.

If $N_f < \frac{11}{2} K$, the one loop calculation  indicates asymptotic
freedom. The two - loop calculations  \cite{Appelquist} indicate that the
chiral symmetry breaking occurs at
\begin{equation}
N_f < N_c \sim K \frac{100 K^2 - 66}{25 K^2 - 15} \sim 4 K \label{NC}
\end{equation}
This is required for the appearance of gauge boson masses.

In our model we have  three generations of Farhi - Susskind technifermions.
Therefore, their number is $24 > 4 N_{TC} = 16$. However, it is important that
only such technifermions enter (\ref{NC}), the masses of which are of the order
of $\Lambda_{TC}$ and smaller ($\Lambda_{TC}$ is the $SU(4)$ analogue of
$\Lambda_{QCD}$). Therefore, we suppose that the masses of the third generation
technifermions and, probably, the masses of some of the second generation
technifermions  are essentially larger, than the Technicolor scale. We also
assume that the masses of the fermions that carry the indices of higher
Hypercolor groups are essentially larger than the Technicolor scale. So, they
do not affect the Technicolor dynamics. Thus  the $SU(4)$ interactions lead to
the chiral symmetry breaking and provide $W$ and $Z$ bosons with their masses.

If the number of fermions approach $N_c \sim 4 N_{TC}$, then the behavior of
the model becomes close to conformal. In this case the effective charge becomes
walking instead of running \cite{walking}. So, in our case (two generations of
fermions for $N_{TC} = 4$) the behavior of the technicolor may be close to
conformal. It may not be conformal if the fermions of the second generation are
also extra massive.

As for the higher Hypercolor groups, already for $SU(5)$ interactions the
number of the first generation hyperfermions (fermions carrying $SU(5)$ index)
is $2(1 + 3 + 4 + 12) = 40 > \frac{55}{2} = 27.5$. We suppose their masses are
close to each other. That's why the Hypercolor forces at $K > 4$ are not
asymptotic free, and do not confine. As a result the Landau pole is present in
their effective charges. This means that our model does not have a rigorous
continuum limit, and should be considered as a finite cutoff model. At the
energies of the order of this cutoff the new theory should appear that
incorporates the Hypercolor tower as an effective low energy theory. In
principle, this scale may be extremely large, even of the order of Plank mass
depending on the value of $g^2_{SU(K)}$ at low energies. Very roughly this
scale (as given by the $SU(5)$ effective charge) can  be estimated as

\begin{equation}
\Lambda_h = e^{\frac{6 \pi}{(2N_f - 55) \alpha_{SU(5)}(1 \, {\rm Tev})}} \,
{\rm Tev}
\end{equation}
Say, if three generations are involved, and $\alpha_{SU(5)}(1 \, {\rm Tev}) =
\frac{1}{300}$, then the Landau pole occurs in the $SU(5)$ gauge coupling at
$\Lambda_h \sim 10^{13}\, {\rm Tev}$.

\section{Conclusions}

In this paper we present the construction of the ulltraviolet completion of the
Standard Model. This completion is organized as a tower of Hypercolor gauge
theories with the common gauge group $G = ... \otimes SU(5)\otimes SU(4)
\otimes SU(3) \otimes SU(2) \otimes U(1)/{\cal Z}$. The fermions of the model
may carry indices from the fundamental representation  of any $SU(N)$ subgroup
of the  gauge group $G$. (Index of each representation may appear only once.)
In addition we require that the $SU(2)$ subgroup acts only on the left - handed
spinors. Then the only uncertainty is the hypercharge assignment. In order to
fix the hypercharges of the fermions we first suppose that there exists a one -
to one correspondence between the left - handed and the right handed fermions.
This correspondence is related to parity conjugation. The definition of the
parity conjugation uses an auxiliary $SU(2)$ field. The unitary gauge can
always be fixed, which gives to the left - handed doublets their conventional
Standard Model form. So, for any set of $SU(N)$ indices there exist two right -
handed fermions and one left - handed fermion. Next, in order to fix their
hypercharges we require that the $Z_6$ symmetry of the Standard Model is
continued to the Hypercolor tower. We choose this continuation in the form
(\ref{symlatWS}). The resulting discrete symmetry $\cal Z$ fixes hypercharge of
each hyperfermion up to an arbitrary integer number. We prove that an
additional constraint may be imposed on these integer numbers such that the
chiral anomaly is absent even if the sequence (\ref{G}) is ended at any rang of
the Hypercolor $SU(N)$ subgroups.

The main reason why we apply the additional $\cal Z$ symmetry to the
construction of the model is that we guess the $Z_6$ symmetry of the Standard
Model cannot appear accidentally, and it should be continued in a certain way
to the more fundamental theory. Of course, our choice (\ref{symlatWS}) is just
one of the possible ways of this continuation. Besides, we may suppose that
their exists the more fundamental theory that has our Hypercolor tower as a low
energy approximation. Then, its monopole content has a deep relation to the
discrete $\cal Z$ symmetry. Namely, all Hypercolor subgroups are involved in
the formation of monopole configurations if the additional $\cal Z$ symmetry is
present.

The dynamics of the theory is organized in such a way that the $SU(4)$
interactions are confining, provide chiral symmetry breaking, and give rise to
the masses of $W$ and $Z$ bosons. In order to provide necessary properties of
the $SU(4)$ interactions we suppose that the third generation technifermions
(and the hyperfermions of higher Hypercolor groups) have masses much larger
than the Technicolor scale $\sim 1$ Tev. The higher Hypercolor interactions are
not confining. The theory admits Landau poles in their effective charges. The
correspondent energy scale may, however, be extremely large, in principle, it
may be even of the order of Plank mass.

The essential feature of our model is that the fermion mass formation is not
related to the transformation of technifermions into the other physical states.
We suppose the fermion masses are generated at the energies much higher than
the Technicolor scale. In order to incorporate fermion masses to the theory we
use the auxiliary scalar $SU(2)$ field. The action does not contain dynamical
term with the  derivatives of this field. The only place in the action, where
this field appears is the fermion mass term. That's why this auxiliary field
does not cause the well known problems of the usual Standard Model Higgs sector
with the dynamical scalar field.

The Hypercolor model described in this paper may be related to the following
picture of fundamental forces at the energies above $1$ Tev. We can consider an
analogy to the Condensed Matter systems, where no detailed description of
microscopic physics is known. Nevertheless, in such systems the simple
excitations and their interactions may be described in an elegant and simple
way. Symmetry properties play an important role in such a description. Our
tower of Hypercolor gauge groups may play the role of such effective
description of an unknown microscopic physics, that is to appear above $1$ Tev.
From this point of view the appearance of all gauge groups $SU(N)$ with any $N$
and all possible fermions from their fundamental representations is quite
natural.

\begin{acknowledgments}
This work was partly supported by RFBR grants  09-02-00338, 08-02-00661, and
07-02-00237, by Grant for leading scientific schools 679.2008.2, by Federal
Program of the Russian Ministry of Industry, Science and Technology No
40.052.1.1.1112.
\end{acknowledgments}

\clearpage


\begin{thebibliography}{99}

\bibitem{Technicolor}
 Christopher T. Hill, Elizabeth H. Simmons,
 Phys.Rept. 381 (2003) 235-402;
Erratum-ibid. 390 (2004) 553-554

Kenneth Lane, hep-ph/0202255

R. Sekhar Chivukula, hep-ph/0011264

\bibitem{ExtendedTechnicolor}
 Thomas Appelquist, Neil Christensen, Maurizio Piai, Robert Shrock, Phys.Rev. D70 (2004) 093010

 Adam Martin, Kenneth Lane, Phys.Rev. D71 (2005) 015011

 Thomas Appelquist, Maurizio Piai, Robert Shrock, Phys.Rev. D69
(2004) 015002

Robert Shrock, hep-ph/0703050

 Adam Martin, Kenneth Lane, Phys.Rev. D71 (2005) 015011

\bibitem{LittleHiggs}
 Maxim Perelstein, Prog.Part.Nucl.Phys.58:247-291,2007

\bibitem{SUSYLHC}
 Jan Kalinowski,
Int.J.Mod.Phys.A22:5920-5934,2007


\bibitem{Extra}
 Thomas Appelquist, Ho-Ung Yee, Phys.Rev. D67 (2003) 055002

\bibitem{TevGrav}
 A. Nicolaidis, N.G. Sanchez,
 Mod.Phys.Lett. A20 (2005) 1203-1208



\bibitem{TEV}
J.A. Casas, J.R. Espinosa, I. Hidalgo, hep-ph/0607279

F. del Aguila, R. Pittau,  Acta Phys.Polon. B35 (2004) 2767-2780

\bibitem{VZ2008}
 A.I.Veselov, M.A.Zubkov, JHEP 0812:109,2008

\bibitem{BVZ2007}
B.L.G.~Bakker, A.I.~Veselov, and M.A.~Zubkov,  J. Phys. G: Nucl. Part. Phys. 36
(2009) 075008.



\bibitem{Appelquist}
 Thomas Appelquist, John Terning, L.C.R. Wijewardhana, Phys.Rev.Lett.
79 (1997) 2767-2770

 Neil D. Christensen, Robert Shrock, Phys.Rev. D72 (2005) 035013

 Masafumi Kurachi, Robert Shrock, Koichi Yamawaki, hep-ph/0704.3481

 T. Appelquist, F. Sannino, Phys.Rev. D59 (1999) 067702

  T. Appelquist, P.S. Rodrigues da Silva, F. Sannino, Phys.Rev. D60 (1999) 116007

\bibitem{minimal_walking}
R. Foadi, M.T. Frandsen, T. A. Ryttov, F. Sannino, arXiv:0706.1696

Sven Bjarke Gudnason, Chris Kouvaris, Francesco Sannino,   Phys.Rev. D73 (2006)
115003

D.D. Dietrich (NBI), F. Sannino (NBI), K. Tuominen, Phys.Rev. D72 (2005) 055001





\bibitem{Sannino_t}
F. Sannino, arXiv:0804.0182

N. Evans and F. Sannino, arXiv:hep-ph/0512080.



\bibitem{FS}
E.Farhi, L.Susskind, Phys.Rev.D 20, 3404, 1979


\bibitem{Z6}
C.Gardner, J.Harvey, Phys. Rev. Lett. {\bf 52} (1984) 879

Tanmay Vachaspati, Phys.Rev.Lett. 76 (1996) 188-191

Hong Liu, Tanmay Vachaspati, Phys.Rev. D56 (1997) 1300-1312

\bibitem{Z6f}
K.S.~Babu, I.~Gogoladze, and K.~Wang, Phys. Lett. B {\bf 570}, 32 (2003);\\
K.S.~Babu, I.~Gogoladze, and K.~Wang, Nucl. Phys. B {\bf 660}, 322 (2003);


\bibitem{BVZ2003}
B.L.G.~Bakker, A.I.~Veselov, and M.A.~Zubkov, Phys. Lett. B {\bf  583}, 379
(2004);

\bibitem{Z2007}
M.A.~Zubkov, Phys. Lett. B {\bf  649}, 91 (2007), Erratum-ibid.B655:91,2007.


\bibitem{Z2007t}
M.A.Zubkov, arXiv:0707.0731, ITEP-LAT/2007-11, Physics Letters B674 (2009),
325-329


\bibitem{PATI}
J.C.Pati, S.Radjpoot, A.Salam, Phys. Rev.  {\bf D 17}, 131 (1978)


\bibitem{Weinberg}
S.Weinberg, "The Quantum Theory of Fields", Cambridge, University press, 2001


\bibitem{Align}
J. Preskill, Nucl. Phys. B177, 21 (1981); M. E. Peskin, Nucl. Phys. B175, 197
(1980).


\bibitem{HooftPolyakov}
Gerard 't Hooft, Nucl.Phys.B79:276-284,1974

Alexander M. Polyakov, JETP Lett.20:194-195,1974, Pisma
Zh.Eksp.Teor.Fiz.20:430-433,1974



\bibitem{PUT}
Andrzej~J.~Buras, P.Q.~Hung, Phys. Rev. D {\bf 68}, 035015 (2003).

Andrzej~J.~Buras, P.Q.~Hung, Ngoc-Khanh~Tran, Anton~Poschenrieder, and
Elmar~Wyszomirski, Nucl.Phys. B {\bf 699}, 253 (2004);

 Mehrdad~Adibzadeh and
P.Q.~Hung, hep-ph/0705.1154.




\bibitem{walking}

Thomas Appelquist, Anuradha Ratnaweera, John Terning, L. C. R. Wijewardhana,
Phys.Rev. D58 (1998) 105017





\end{thebibliography}
\end{document}